# On the calculation of the quality factor in contemporary photonic resonant structures

**Thomas Christopoulos,[1,*] Odysseas Tsilipakos,[2,#] Georgios Sinatkas,[1] and Emmanouil E. Kriezis[1]**

[1] *School of Electrical and Computer Engineering, Aristotle University of Thessaloniki (AUTH), 54124 Thessaloniki, Greece*
[2] *Institute of Electronic Structure and Laser, Foundation for Research and Technology Hellas (FORTH), 71110 Heraklion, Crete, Greece*
**cthomasa@ece.auth.gr*
*#otsilipakos@iesl.forth.gr*

**Abstract:** The correct numerical calculation of the resonance characteristics and, principally, the quality factor $Q$ of contemporary photonic and plasmonic resonant systems is of utmost importance, since $Q$ defines the bandwidth and affects nonlinear and spontaneous emission processes. Here, we comparatively assess the commonly used methods for calculating $Q$ using spectral simulations with commercially available, general-purpose software. We study the applicability range of these methods through judiciously selected examples covering different material systems and frequency regimes from the far-infrared to the visible. We take care in highlighting the underlying physical and numerical reasons limiting the applicability of each one. Our findings demonstrate that in contemporary systems (plasmonics, 2D materials) $Q$ calculation is not trivial, mainly due to the physical complication of strong material dispersion and light leakage. Our work can act as a reference for the mindful and accurate calculation of the quality factor and can serve as a handbook for its evaluation in guided-wave and free-space photonic and plasmonic resonant systems.

## 1. Introduction

In the past decades, optical micro- and nano-resonators have been established as indispensable components for guided-wave and free-space photonic and plasmonic systems due to their ability to trap light, provide frequency selective response and foster enhanced local fields leading to strong light-matter interaction [1]. They have been realized in almost all available material systems, ranging from silicon [2,3], III-V semiconductors [4], fused silica [5], and chalcogenide glasses [6], to the emerging field of 2D photonic materials [7]. Resonant systems come in many forms, namely, ring/disk and sphere cavities [2, 5], (plasmonic) nanoantennas [8], photonic crystal structures [9, 10], and metamaterials/metasurfaces [11, 12], to name a few. They can be exploited for a variety of applications capitalizing on their strong spatial field confinement (even below the diffraction limit) and long temporal energy storage. A glimpse into the resonant systems "galaxy" is shown in Fig. 1, including guided-wave and free-space systems for a broad range of frequency regimes from the THz/far-infrared (FIR) to the near-infrared (NIR) and visible.

Because of the numerous applications of resonant systems, the accurate calculation of their resonance characteristics, notably the resonance frequency and the quality factor ($Q$-factor), is of significant importance. In particular, the quality factor, besides directly specifying the energy decay rate and spectral bandwidth, plays an important role in determining the efficiency of nonlinear phenomena [13] and the modification of spontaneous emission rates of quantum emitters in the vicinity of resonant structures [14]. In addition, it is frequently used to feed temporal coupled-mode theory (CMT) models [15, 16] that can efficiently assess the response of linear and, importantly, complex nonlinear resonant systems [17–19], otherwise requiring cumbersome nonlinear full-wave simulations.

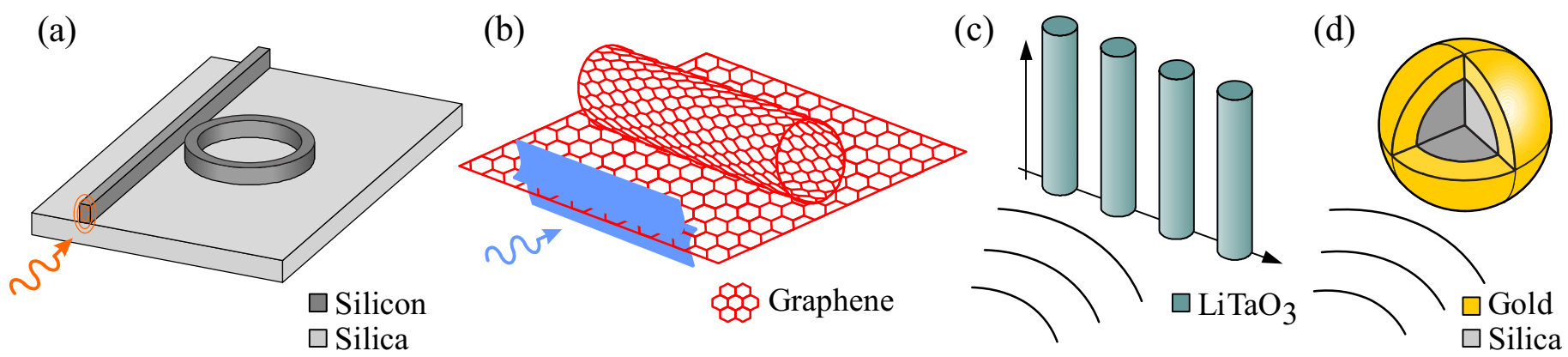

Fig. 1. Characteristic contemporary guided-wave and free-space resonant photonic/plasmonic systems. (a) Silicon microring resonator coupled to an access waveguide. (b) Carbon tube resonator accessed through a graphene sheet. (c) Free-space dielectric rod metasurface. (d) Free-space plasmonic core-shell nanoparticle.

Thanks to the wide availability of simulation tools for solving the Maxwell's equations, it has been made possible to numerically calculate quality factors in geometrically complex resonators. However, the correct estimation of the resonance characteristics in contemporary photonic, plasmonic, and 2D-material-comprising structures requires extra attention because of complications arising from the presence of strong material dispersion, considerable ohmic loss, and significant radiation leakage. Such issues are frequently overlooked or misconceived in the literature when calculating the $Q$-factor. In this work, we systematically organize and report the available routes for calculating the quality factor using commercially available, general purpose computational electromagnetic tools. Through judiciously selected examples that gradually introduce physical complications (i.e., strong ohmic loss, radiation leakage, and material dispersion), we highlight the applicability range of each available method. Note that customized techniques to accurately specify the complex resonance frequency (and therefore the quality factor) in cases of strong material dispersion and/or radiation leakage have been reported in the literature [14, 20–22] providing physical understanding and insight; however, they require specialized numerical approaches. In sharp contrast, this work focuses on straightforward calculation methods that utilize conventional, widely available, electromagnetic software tools.

The remainder of the paper is organized as follows: In Sec. 2 we briefly present the available methods for calculating the quality factor and verify their equivalence. Subsequently, we apply the aforementioned methods in four judiciously chosen examples, characteristic of contemporary photonic/plasmonic platforms, covering a wide spectral range from the far-infrared to the visible. Specifically, we study (i) a silicon ring resonator in the near-infrared (Sec. 3), (ii) a graphene plasmonic tube resonator in the far-infrared (Sec. 4), (iii) a polaritonic rod meta-atom and the respective metasurface in the far-infrared (Sec. 5), and (iv) a plasmonic core-shell nanoparticle in the visible (Sec. 6). By using the commercial software COMSOL Multiphysics, based on the frequency-domain finite-element method (FEM), we highlight alternative routes to correctly calculating the resonance frequency and quality factor in each case, additionally validating the range of their applicability through meticulous CMT studies. The conclusions are provided in Sec. 7.

## 2. Methods for calculating the quality factor

In a resonant structure, the quality factor is a measure of the overall loss experienced by the system; lower loss corresponds to higher values of the $Q$-factor, improving the ability of the resonator to trap light. All kinds of possible losses may lower $Q$: ohmic loss (conductor and/or dielectric loss), radiation loss, and coupling loss due to external feeding channels such as input/output waveguides. Various methods are reported in the literature on calculating $Q$. Here, we briefly review them and show that they are, in principle, equivalent. Subsequently, we will highlight the range of their applicability, using four characteristic examples of gradually increasing physical complexity.

### 2.1. Definition based on stored energy: The "Eigenmode" and "Field distribution" methods for calculating $Q$

The quality factor is most commonly defined as the ratio of the stored energy in the resonator over the dissipated energy per optical cycle [23],

$$Q = \omega_0 \frac{W}{P_{\text{loss}}}, \tag{1}$$

where $\omega_0$ is the (angular) resonance frequency, $W$ is the stored energy, and $P_{\text{loss}}$ represents one or more loss mechanisms. Depending on the nature of $P_{\text{loss}}$, various $Q$-factors emerge: (i) the *resistive quality factor*, $Q_{\text{res}}$, associated with ohmic (resistive) loss $P_{\text{res}}$, (ii) the *radiation quality factor*, $Q_{\text{rad}}$, associated with radiation loss $P_{\text{rad}}$, and (iii) the *external or coupling quality factor*, $Q_e$, associated with the coupling loss $P_e$ due to external feeding channels (e.g., input/output waveguides). Furthermore, it is quite common to group resistive and radiation losses ($P_i = P_{\text{res}} + P_{\text{rad}}$) under the *intrinsic or unloaded quality factor*, $Q_i$, to fully describe an isolated resonator. Finally, in the presence of coupling to external waveguides, the *loaded quality factor*, $Q_\ell$, is defined as the sum of resistive, radiation, and coupling losses ($P_\ell = P_{\text{res}} + P_{\text{rad}} + P_e$). Based on Eq. (1), the $Q$-factors described above are related through

$$\frac{1}{Q_\ell} = \frac{1}{Q_i} + \frac{1}{Q_e} = \frac{1}{Q_{\text{res}}} + \frac{1}{Q_{\text{rad}}} + \frac{1}{Q_e}. \tag{2}$$

Obviously, Eq. (2) holds only when the individual quality factors are consistently calculated.

The $Q$-factor definition can be evaluated using either the eigenmode (mode profile) of the resonant system, obtained after solving for the respective eigenvalue problem ("*eigenmode*" method for calculating $Q$), or utilizing the field distribution (full or scattered field), obtained from a driven (i.e., with excitation) time-harmonic simulation ("*field-distribution*" method for calculating $Q$). In general, their results may differ; we will systematically examine both methods and highlight the range of their applicability using carefully chosen examples.

### 2.2. The "eigenfrequency" method for calculating $Q$

In an alternative approach, the quality factor may be calculated using the complex eigenfrequency $\tilde{\omega} = \omega' + j\omega''$ ($\omega'$, $\omega'' > 0$ for the adopted herein $\exp\{+j\omega t\}$ time-harmonic convention) that characterizes a resonance mode (usually referred to as a quasi-normal mode, QNM) through

$$Q = \frac{\omega'}{2\omega''}. \tag{3}$$

The real part of the complex eigenfrequency $\omega' = \omega_0$ is the resonance frequency and the imaginary part $\omega'' = 1/\tau$ encompasses every loss mechanism and is inversely proportional to the photon lifetime in the cavity, $\tau$. This eigenfrequency is the eigenvalue of the respective eigenproblem. By applying the Poynting theorem on a volume $V$ (enclosed by the surface boundary $\partial V$) [24], Eq. (4) is derived,

$$\begin{aligned} \iiint_V \text{Re}\{\nabla \cdot \mathbf{S_c}\} \text{d}V = \oint\!\!\!\oint_{\partial V} \text{Re}\{\mathbf{S_c}\} \cdot \hat{\mathbf{n}} \, \text{d}S =& 2\omega'' \iiint_V \frac{1}{4} \left( \varepsilon_0 \varepsilon_r' |\mathbf{E}|^2 + \mu_0 |\mathbf{H}|^2 \right) \text{d}V \\ & - \iiint_V \frac{1}{2} \omega' \varepsilon_0 \varepsilon_r'' |\mathbf{E}|^2 \text{d}V - \iiint_V \frac{1}{2} \sigma |\mathbf{E}|^2 \text{d}V, \end{aligned} \tag{4}$$

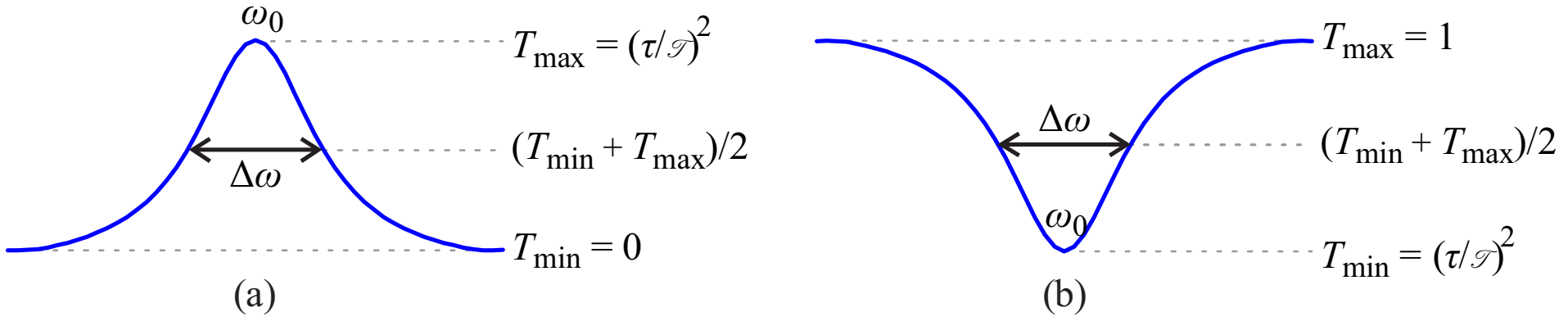

Fig. 2. (a) Lorentzian lineshape and (b) inverse Lorentzian lineshape. Minimum/maximum values and the FWHM bandwidth are clearly marked.

with $\varepsilon_r = \varepsilon_r' - j\varepsilon_r''$ being the complex permittivity of lossy dielectrics, $\sigma$ being the real conductivity, and $\mu_r = 1$. When material dispersion is absent, the terms in the right hand side of Eq. (4) amount to the stored energy and the resistive power loss, taking the condensed form

$$P_{\text{rad}} + P_e = 2\omega'' W - P_{\text{res}}. \tag{5}$$

In the forthcoming Secs. 3-4, we will reestablish the dispersive expression of the stored energy [24, 25], leading to the breakdown of Eq. (5). Alternatively, Eq. (5) can be easily retrieved by noting that the temporal energy decay due to losses is proportional to $\exp\{-2\omega'' t\}$ and, thus, $P_{\text{loss}} = -\mathrm{d}W/\mathrm{d}t = 2\omega'' W$. By substituting Eq. (5) to the $Q$-factor definition of Eq. (1), Eq. (3) is easily recovered, verifying the equivalence of the calculation methods.

It must be noted that the complex nature of $\tilde{\omega}$, ensuring the mode decay in time, leads to an exponential divergence of the eigenmode in space with the radial dimension $r$, a phenomenon that has been also observed in early works on leaky cavities [1, 14, 26]. Specifically, it is easily seen that $\exp\{+j\tilde{\omega}(t - r/c)\}$ decays as $t \to \infty$ ($\exp\{-t/\tau\}$) but diverges as $r \to \infty$ ($\exp\{+r/\tau c\}$). Lossy cavities have also complex eigenfrequencies but the exponential divergence is not noticed unless the radiation leakage is significant.

From a slightly different point of view, a photonic resonator may be described as a lumped harmonic oscillator which, under some approximations, can be modeled using a first-order ordinary differential equation [15]

$$\frac{\mathrm{d}a(t)}{\mathrm{d}t} = \left(j\omega_0 - \frac{1}{\tau}\right) a(t), \tag{6}$$

where $a(t)$ is the amplitude of the cavity mode normalized so that $|a|^2 \equiv W$. This framework is usually referred to as the temporal coupled-mode theory. In this context, power dissipation of the cavity mode is given by $P_{\text{loss}} = -\mathrm{d}|a|^2/\mathrm{d}t = -\mathrm{d}W/\mathrm{d}t = (2/\tau)W$. Substituting in Eq. (1), returns $Q = \omega_0\tau/2$. On the other hand, the complex eigenfrequency $\tilde{\omega} = \omega_0 + j/\tau$ of Eq. (6) results in the same $Q$ when inserted in Eq. (3), revealing the equivalence with the already discussed approaches.

In conclusion, Eq. (3) offers a different route to calculating the quality factor, termed as the "*eigenfrequency*" method for calculating $Q$. Depending on the specific properties of the resonator, Eq. (3) can be applied to specify some of the aforementioned $Q$-factors. The applicability range will be elucidated with the following examples.

### 2.3. The "spectral response" method for calculating $Q$

Finally, a third approach to calculate the $Q$-factor, naturally suited for application to experimental measurements, is based on exploiting the clear Lorentzian lineshape (Fig. 2) that a single leaky/lossy resonance exhibits [27] (in the transmission/reflection power coefficient, the absorption/scattering cross-section, etc.), obtained alternatively after driven time-harmonic simulations. We term this the "*spectral response*" method for calculating $Q$. The $Q$-factor is given by

$$Q = \frac{\omega_0}{\Delta\omega}, \tag{7}$$

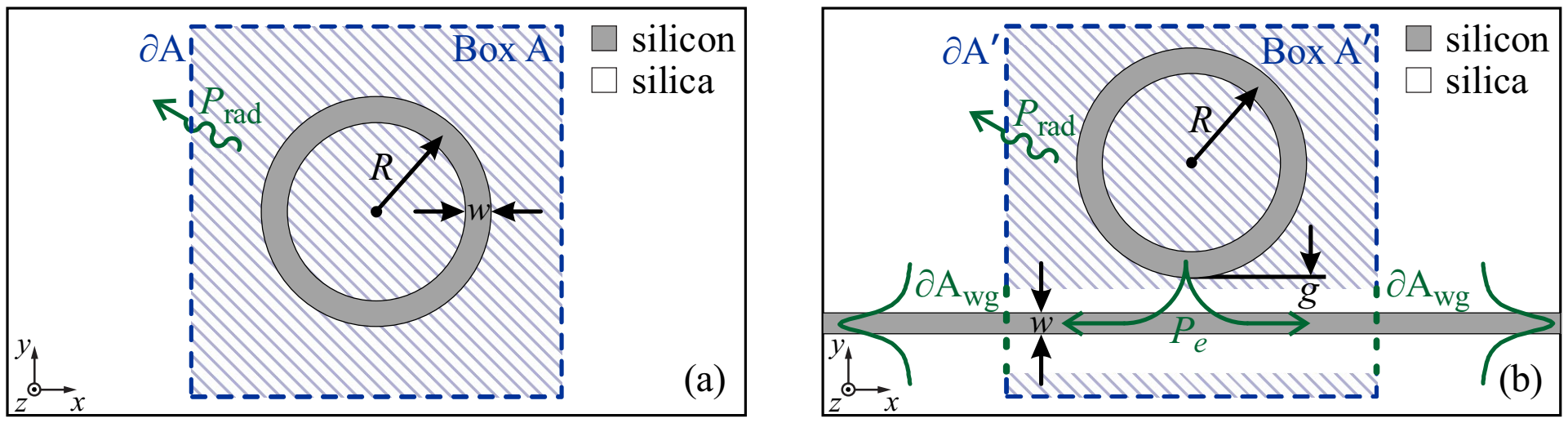

Fig. 3. Schematic of (a) an uncoupled and (b) a coupled, silica-clad, silicon ring resonator of radius $R$ with a coupling gap $g$. Energy integration domains and the respective power flux integration boundaries for the correct application of Eq. (1) are clearly highlighted.

where $\Delta\omega$ is the full-width at half-maximum (FWHM). Starting from the complex eigenfreqency $\tilde{\omega}$, the mode decay in the time domain is proportional to $\exp\{-t/\tau\}\exp\{j\omega_0 t\}$ for $t > 0$. In the frequency domain, its squared norm takes the Lorentzian form $1/[(1/\tau)^2 + (\omega - \omega_0)^2]$, Fig. 2(a). The FWHM at $(T_{\min} + T_{\max})/2$ is easily calculated being $\Delta\omega = 2/\tau$, resulting in $Q = \omega_0\tau/2$, fully consistent with Eq. (3).

Similarly, in the CMT context, the Lorentzian function of a single resonance takes the form $(1/\mathscr{T})^2/[(1/\tau)^2 + (\omega - \omega_0)^2]$, where $\mathscr{T}$ is a case-dependent lifetime (part of the total $\tau$) related to the nature of the physical system and the coupling conditions. Again, the FWHM is found equal to $\Delta\omega = 2/\tau$; thus, Eq. (7) results in the same $Q$ value as Eqs. (1) and (3). Resonances with inverse Lorentzian lineshape [Fig. 2(b)], modeled by $[(1/\mathscr{T})^2 + (\omega - \omega_0)^2]/[(1/\tau)^2 + (\omega - \omega_0)^2]$, similarly result in the same $\Delta\omega = 2/\tau$ and, as a consequence, the same quality factor expression.

## 3. Silicon ring resonator at the NIR

Towards the evaluation of the $Q$-factor calculation methods presented in Sec. 2, we consider hereafter resonant systems with different characteristics and gradually increasing complexity to highlight each method's applicability range. First, we examine a silicon microring resonator in the NIR. Such structures are widely used as filtering elements in telecom applications. The structure under study is depicted in Fig 3(a) and is the 2D equivalent of Fig. 1(a). Examining a 2D example allows for achieving excellent numerical accuracy and thus reaching sound conclusions, without sacrificing any of the underlying physical considerations. The ring radius $R$ is in the order of 1 $\mu$m and the slab width is $w = 200$ nm. A bus waveguide of the same width is placed near the ring at an edge-to-edge distance $g$ in the order of 300 nm to couple it with the outside world, Fig 3(b). The ring is designed to resonate near 1.55 $\mu$m, where Si has a refractive index of $n_{\rm Si} = 3.478$. The cladding is $SiO_2$ with $n_{SiO_2} = 1.45$. For both silicon and silica, refractive index dispersion is taken into account via the appropriate Sellmeier equation [28, 29].

First, we examine the uncoupled ring in order to calculate the resonance wavelength and intrinsic quality factor. Both silicon and silica are assumed to exhibit negligible ohmic losses around 1.55 $\mu$m; $Q_{\rm res} \to \infty$ and thus $Q_i$ coincides with $Q_{\rm rad}$. FEM eigenvalue simulations are conducted by solving the curl-curl electric field vector wave equation for various radii, corresponding to modes of different azimuthal order, all with resonance wavelengths around 1.55 $\mu$m. As already mentioned, the real part of the calculated complex eigenfrequency $\tilde{\omega}$ represents the resonance frequency $\omega_0$. A single-step solution scheme is not adequate for correctly estimating $\omega_0$ due to material dispersion. An iterative approach is usually adopted by refreshing material properties based on the $\omega_0$ obtained in the previous step (or in some cases a weighted average among previous steps), until convergence. In this particular example, dispersion is weak, requiring only a few iterations.

The intrinsic $Q$-factor can be easily calculated through the "eigenfrequency" calculation method [Eq. (3)]. Although in specifying $\omega_0$ we were able to take dispersion into account with the iterative procedure described, this method is not capable of correctly calculating $Q$ in dispersive systems, as it will become evident in the next examples, failing to physically encapsulate the dispersion of the underlying material properties. Nevertheless, as already mentioned, in this particular system dispersion is weak. The results for five consecutive modes ($m = 9$ up to $m = 13$) are presented as black circles in Fig. 4(a). Greater radii result in lower radiation loss or, equivalently, higher $Q$-factors, as expected.

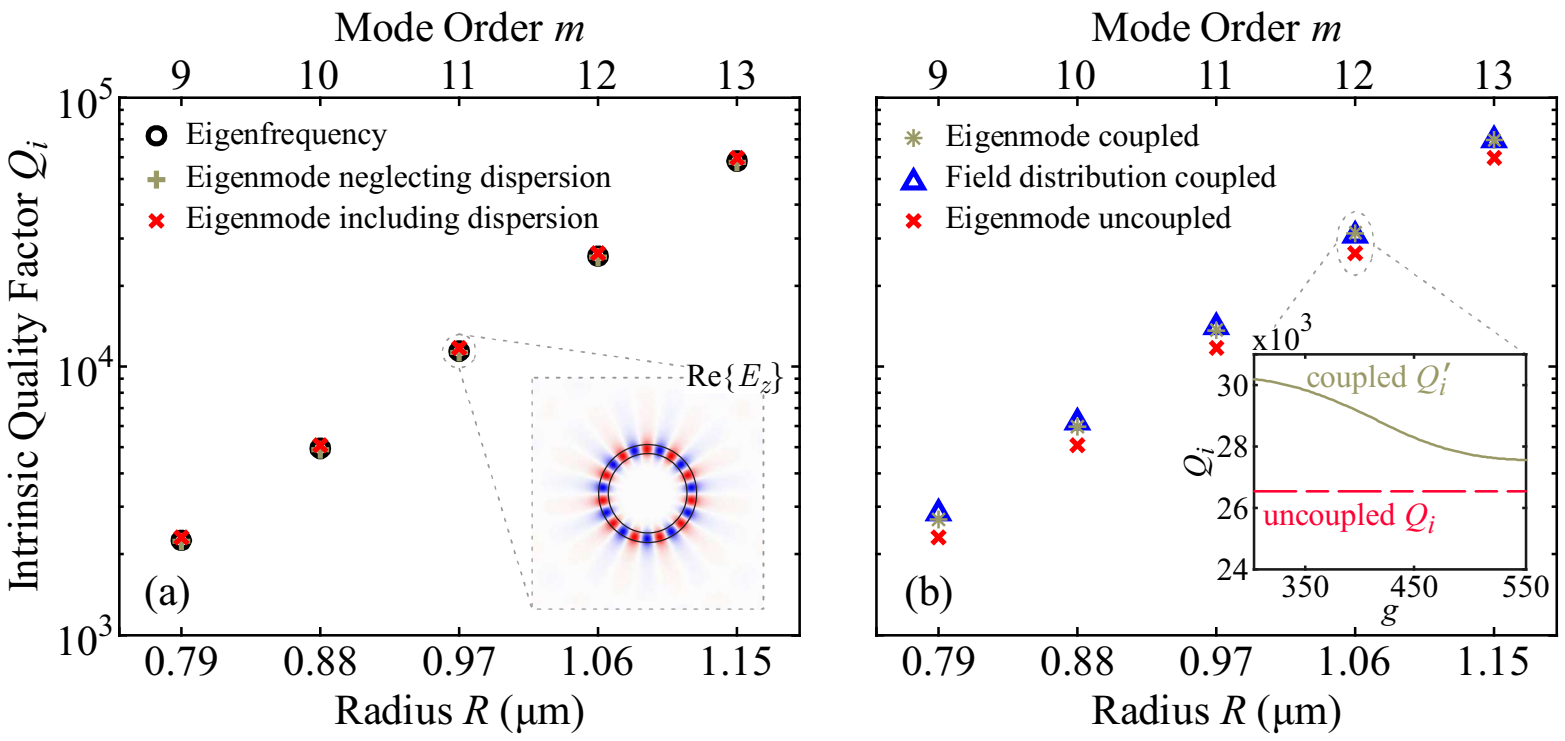

Fig. 4. (a) Uncoupled intrinsic $Q$-factor calculated using the eigenfrequency (black circles) and the respective eigenmode (yellow and red crosses). (b) Coupled intrinsic $Q$-factor (yellow stars and blue triangles). External coupling results in higher $Q_i'$ since a non-negligible part of the radiation is coupled to the bus waveguide (inset). All modes have resonance wavelengths around 1.55 $\mu$m.

Alternatively, the respective *eigenmode*, i.e., the electromagnetic mode profile, can be used to calculate $Q_i$ by applying Eq. (1). Mode energy $W$ is calculated through

$$W = \frac{1}{4} \iint_A \varepsilon_0 \left. \frac{\partial\{\omega\varepsilon_r(\omega)\}}{\partial\omega} \right|_{\omega=\omega_0} |\mathbf{E}|^2 \mathrm{d}S + \frac{1}{4} \iint_A \mu_0 |\mathbf{H}|^2 \mathrm{d}S, \tag{8}$$

in a box sufficiently large [e.g., the cross-hatched Box A, Fig 3(a)] in order to completely include the tailing edges of the resonance mode. Radiated power $P_{\mathrm{rad}}$ exiting the box is calculated from the outward component of the Poynting vector at the box boundaries, i.e.,

$$P_{\mathrm{rad}} = \int_{\partial A} \mathrm{Re}\{\mathbf{S_c}\} \cdot \hat{\mathbf{n}}\, \mathrm{d}\ell = \frac{1}{2} \int_{\partial A} \mathrm{Re}\left\{\mathbf{E} \times \mathbf{H}^*\right\} \cdot \hat{\mathbf{n}}\, \mathrm{d}\ell. \tag{9}$$

Although the mode exponentially diverges in space, as mentioned in Sec. 2, the energy balance $W/P_{\mathrm{loss}}$ is constant regardless of the integration domain dimensions, Eq. (5). This might seem counter-intuitive at first glance. As the integration domain increases, the stored energy increases with it. Due to the exponential increase of the radiation leakage, the power flux from the boundary increases as well (it otherwise would not). This happens in such a way that $W/P_{\mathrm{loss}}$ remains constant, guaranteeing the consistency between the eigenvector and the respective eigenvalue of the solved eigenproblem. $Q$-factors calculated using this approach are depicted in Fig. 4(a), including (red crosses) or ignoring (yellow crosses) material dispersion, confirming that it is rather weak here. The obtained results also coincide, as expected, with those of the "eigenfrequency" method.

The "eigenmode" method can also be applied to calculate the *intrinsic* $Q$-factor in a *coupled* system, after correctly excluding the energy involved with the bus waveguide and the power exiting the integration domain via the waveguide ports [thus utilizing the cross-hatched Box A′ and its blue dashed limits $\partial$A′, Fig 3(b), to perform the integrations]. In fact, the intrinsic quality factor under coupling, $Q_i'$, is expected to be different from the intrinsic $Q$-factor of the uncoupled resonator $Q_i$. $Q_i'$ depends on the coupling conditions, i.e., it is a function of the coupling gap $g$. This is because a part of the radiated power in the case of the isolated resonator is now evanescently coupled to the bus waveguide, instead of counting as radiation loss. In that sense, $Q_i'(g)$ should be greater than $Q_i$, approaching asymptotically the latter as $g$ becomes greater and the coupling strength weakens. This is indeed verified in Fig. 4(b), where $Q_i'(g)$ is also calculated through the "field distribution" method, using a driven time-harmonic simulation at $\omega_0$ (blue triangles). "Eigenmode" and "field distribution" methods for $Q_i'(g)$ coincide for each gap, as anticipated.

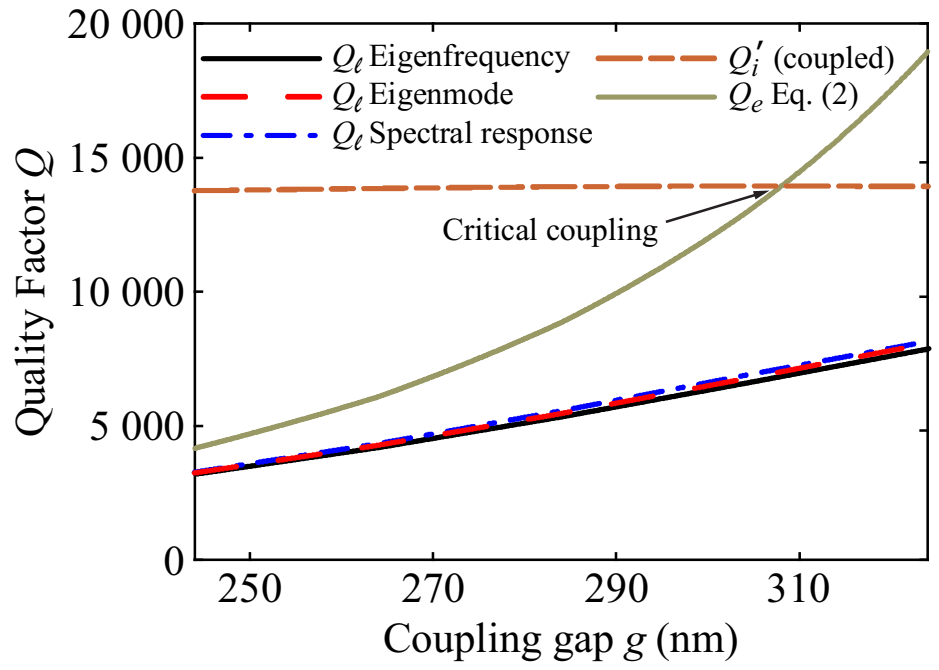

Fig. 5. Loaded $Q$-factor calculated for the $m = 11$ mode using the eigenfrequency (black solid line), the eigenmode (red dashed line), and the spectral response (blue dash-dot line), revealing very good agreement. $Q_i'$ and $Q_e$ are also included, equated at critical coupling.

In order to calculate the loaded quality factor $Q_\ell$, one can again integrate the eigenmode inside Box A′, Fig. 3(b), i.e., apply the "eigenmode" method. To calculate the power coupled to the waveguide, $P_e$, the outward Poynting vector should be integrated over the two green dotted lines [$\partial \mathrm{A}_{\mathrm{wg}}$, Fig. 3(b)], having appropriate length so as to accommodate the extent of the waveguide mode and at least 95% of the guided power. The "field distribution" method can also be applied, howbeit taking special care for disentangling the direct coupling of the incoming wave to the output and the indirect coupling of the radiated wave from the resonator to the output. To do so, Eq. (1) must be applied to the scattered field $\mathbf{E}^{\mathrm{sc}} = \mathbf{E}^{\mathrm{tot}} - \mathbf{E}^{\mathrm{wg}}$, i.e., after omitting the incident field that propagates in the feeding waveguide in the absence of the cavity to achieve the aforementioned disentanglement. Alternatively, the "eigenfrequency" method can be used for calculating $Q_\ell$, since in this particular example dispersion is almost negligible (introducing though minor discrepancies, cf. black solid and red dashed/blue dashed-dotted lines in Fig. 5). Finally, using time-harmonic simulations one can obtain the *spectral response* of the resonator which is expected to be of Lorentzian shape (when resonances are well separated in frequency), calculate FWHM, and then apply Eq. (7). All the aforementioned approaches almost coincide for several coupling gaps $g$, as seen in Fig. 5. In the same figure we have included $Q_i'$ [calculated in A′/∂A′, Fig. 3(b)] and $Q_e$ [calculated through Eq. (2)], useful for identifying critical coupling ($Q_i' = Q_e$) in traveling wave resonators. Note here that $Q_i'$ depends on the coupling gap, as expected, but this trend is not clearly visible because of the axis scaling.

For validating the $Q$-factors calculated throughout this section, we compare the spectral response obtained using full-wave time-harmonic simulations with the one predicted by the CMT-based transmission equation [30]

$$T = \frac{\delta^2 + (1 - r_Q)^2}{\delta^2 + (1 + r_Q)^2}, \tag{10}$$

where $\delta = 2Q_i'(\omega - \omega_0)/\omega_0$ and $r_Q = Q_i'/Q_e$. The results for the most leaky mode ($m = 9$) are illustrated in Fig. 6 for three different coupling gaps, resulting in overcoupling ($g$ = 190 nm), critical coupling ($g$ = 230 nm), and undercoupling ($g$ = 270 nm) conditions, respectively. For comparison, Eq. (10) is also applied using the uncoupled $Q_i$ (red thin curve). As expected, FEM and CMT coincide when $Q_i'$ is used, indicating that this $Q$-factor actually characterizes leaky resonators even under relatively weak coupling conditions. Based on the above, we gather in Table 1 the range of applicability for each $Q$-factor calculation method, with reference to systems where radiation loss may be significant but dispersion is weak and resistive loss negligible, i.e., in most of the commonly considered photonic platforms (silicon-on-insulator [31], fused silica [32], chalcogenide glasses [33], etc.). As can be seen, in such cases all methods are capable of correctly providing the quality factor.

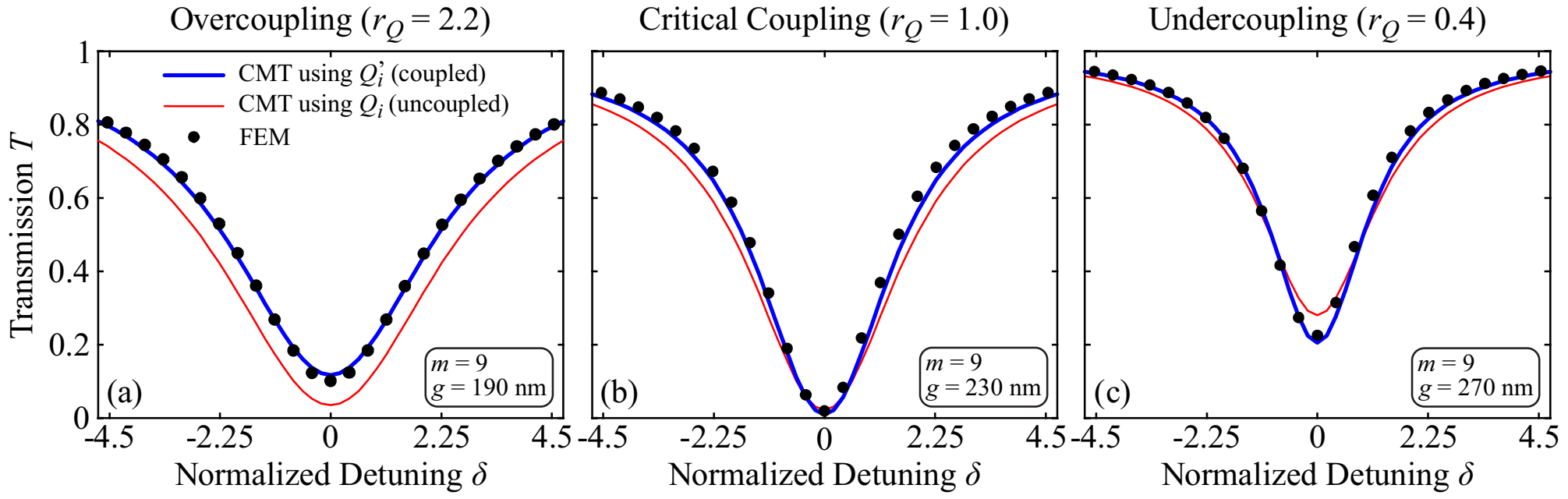

Fig. 6. FEM-based spectral response (black dots) compared to CMT transmission using the coupled $Q_i'$ (blue thick curve) and the uncoupled $Q_i$ (red thin curve). When $Q_i'$ is used, FEM and CMT coincide.

**Table 1. $Q$-factor Calculation Methods in Guided-Wave Systems with Weak Dispersion, Negligible Resistive, and Potentially Significant Radiation Loss. n/a stands for "not applicable."**

| | Eigenproblem | | Time-harmonic (waveguide fed) | |
|---|---|---|---|---|
| Method | Eigenfrequency | Eigenmode | Field distribution | Spectral response |
| $Q_i$ (uncoupled) | ✓ | ✓ | n/a | n/a |
| $Q_i'$ (coupled) | n/a | ✓ | ✓ | n/a |
| $Q_e$ | ✓ using Eq. (2) | ✓ | ✓ using $\mathbf{E}^{\mathrm{sc}}$ | n/a |
| $Q_\ell$ | ✓ | ✓ | ✓ using $\mathbf{E}^{\mathrm{sc}}$ | ✓ |

## 4. Graphene plasmonic tube resonator at THz frequencies

As a second example, we examine a highly dispersive system with significant ohmic loss but low radiation. For diversity, we choose a system that operates in a different frequency band, the far-infrared or THz regime, and supports surface-plasmon polaritons (SPPs); materials that support SPPs are indeed highly dispersive and lossy. Specifically, an infinite graphene-tube resonator of radius $R$ in the order of 2 $\mu$m is considered [Fig. 1(b)], with a resonance frequency around 10 THz. An infinite graphene sheet, acting as the access waveguide, is placed in a distance $g$ of about 1.5 $\mu$m [Fig. 7(b)]. Graphene is computationally modeled as an infinitesimally-thin layer (surface current boundary condition approach), exhibiting a Drude-like surface conductivity $\sigma_{\mathrm{gr}}$ owing to the intraband absorption mechanism that dominates for low energy photons. 2D simulations are conducted for the demonstrations of this section, while graphene conductivity is obtained using the simplified Kubo formula [34] $\sigma_{\mathrm{gr}} = -je^2\mu_c/[\pi\hbar^2(\omega - j/\tau)]$, at a Fermi level of $\mu_c = 0.3$ eV with $\tau = 40$ ps.

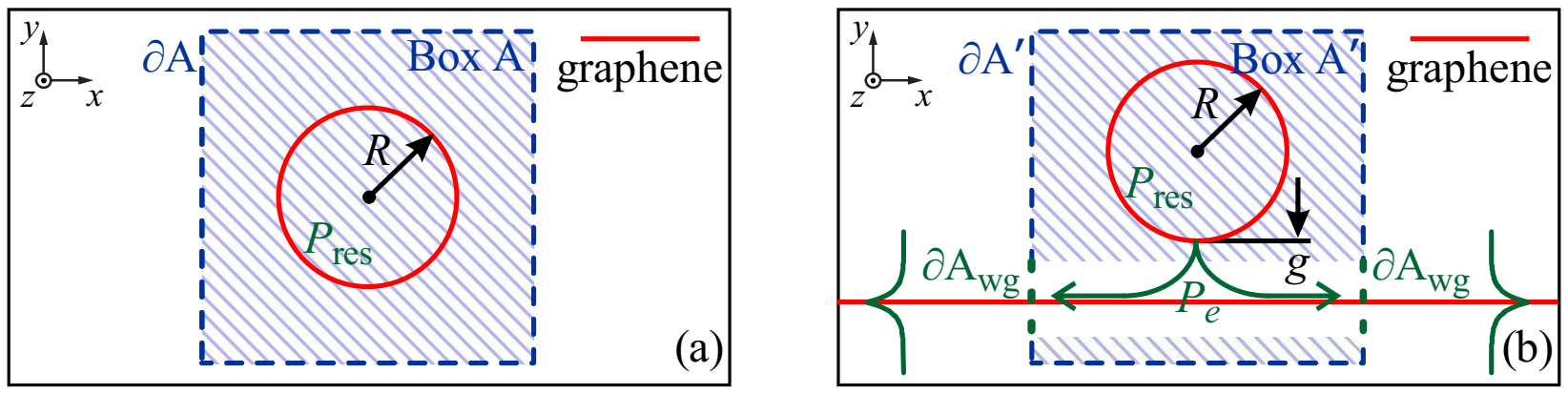

Fig. 7. Schematic of (a) an uncoupled and (b) a coupled graphene tube resonator of radius $R$ with coupling gap $g$. Energy integration domains and the respective power flux integration boundaries for the correct application of Eq. (1) are denoted as well.

The isolated resonator [Fig. 7(a)] is initially examined. Its resonance frequency is easily calculated solving for the respective eigenproblem using the iterative approach introduced in Sec. 3. Regarding the intrinsic $Q$-factor, the results for five consecutive modes ($m = 2$ to $m = 6$) are depicted in Fig. 8(a), using the "eigenfrequency" method. As suggested in the previous section, a typical implementation of an eigenproblem is not capable of capturing the dispersive nature of the underlying materials and, thus, leads to incorrect calculation of the quality factor. An alternative implementation for materials with Drude and Drude-Lorentz dispersion, allowing for the accurate estimation of the complex eigenfrequency, has been proposed [20, 35] but it has not yet been incorporated in any commercial FEM software. In any case, if dispersion is not taken into account the "eigenfrequency" method results coincide with those obtained by the "eigenmode" method (yellow crosses). Surprisingly, coupled $Q_i'$ (red crosses), calculated using the cross-hatched Box A′ with $g = 1.7$ $\mu$m [Fig. 7(b)] also coincide with uncoupled $Q_i$. This can be better understood by separately calculating $Q_{\mathrm{rad}}$ and $Q_{\mathrm{res}}$. Using the "eigenfrequency" method, $Q_{\mathrm{rad}}$ is easily retrieved after setting $\mathrm{Re}\{\sigma_{\mathrm{gr}}\} = 0$, with $Q_{\mathrm{res}}$ subsequently resulting from Eq. (2). For the "eigenmode" method, on the other hand, $P_{\mathrm{rad}}$ and $P'_{\mathrm{rad}}$ should be calculated from Eq. (9) using the integration limits $\partial$A and $\partial$A′, respectively. Finally, resistive loss is given by

$$P_{\mathrm{res}} = \frac{1}{2}\int_{\mathrm{gr}} \mathrm{Re}\left\{\mathbf{E}\cdot\mathbf{J}^*\right\} \mathrm{d}\ell = \frac{1}{2}\int_{\mathrm{gr}} \mathrm{Re}\{\sigma_{\mathrm{gr}}\}|\mathbf{E}_{\parallel}|^2 \mathrm{d}\ell, \tag{11}$$

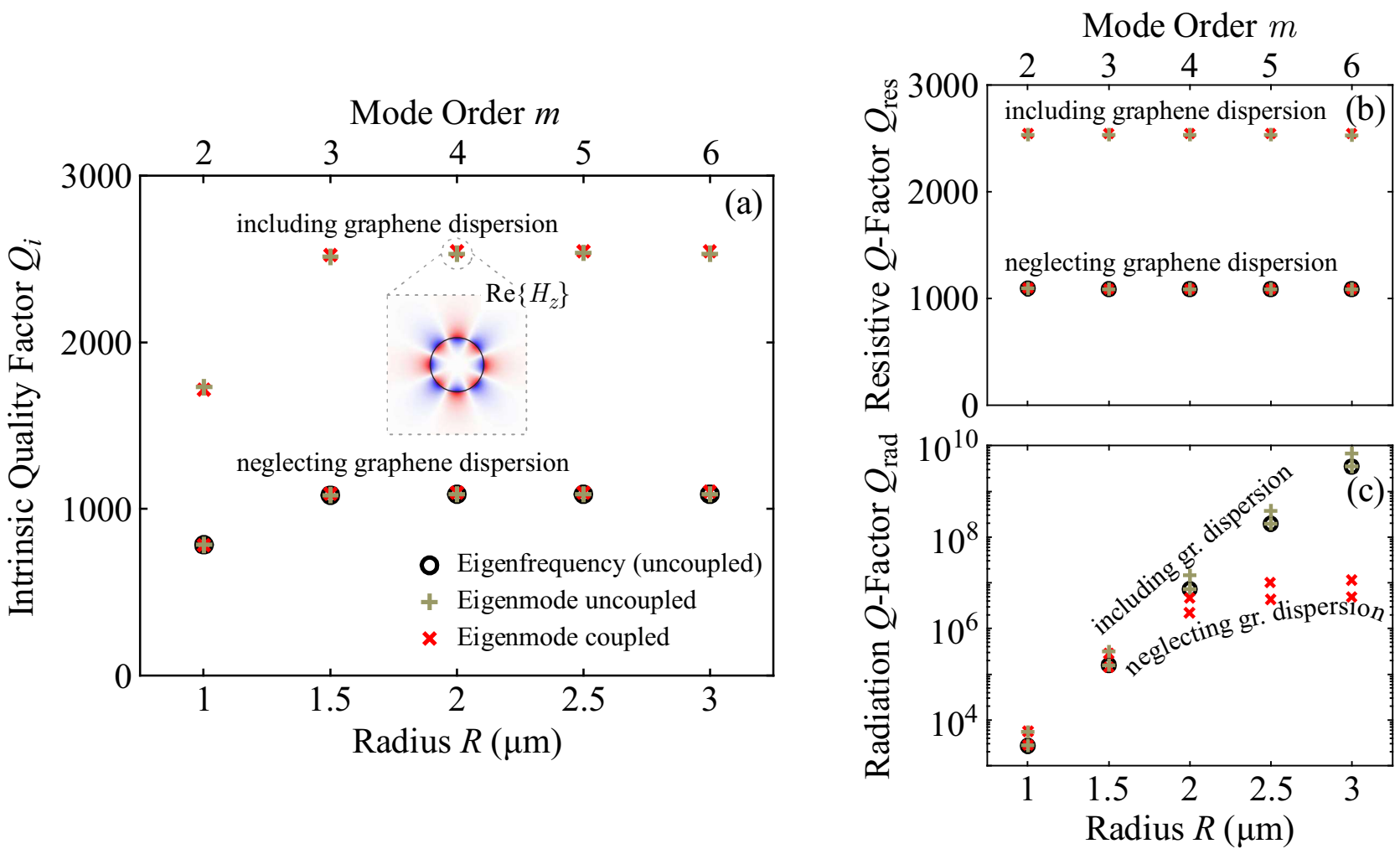

Fig. 8. (a) Intrinsic $Q$-factor calculated after including (upper markers) or neglecting (lower markers) graphene dispersion. The strong impact of dispersion is seen. (b) Resistive and (c) radiation $Q$ for the same conditions. All modes have resonance frequencies around 10 THz.

where "gr" stands for graphene and $\mathbf{E}_{\parallel}$ are the tangential to the graphene sheet $E$-field components. The results, depicted in Figs. 8(b) and 8(c), indicate that the resistive loss dominates over radiation; although $Q_{\mathrm{rad}}$ differs from $Q'_{\mathrm{rad}}$ [Fig. 8(c)], this is not reflected in the respective $Q_i$ and $Q'_i$. Likewise, the absence in practice of any radiation loss keeps $W/P_{\mathrm{res}}$ constant regardless of the integration domain, even though in principle only $W/P_{\mathrm{loss}}$ remains strictly unaffected.

Despite the agreement among the $Q$-factor calculation methods when dispersion is forcedly neglected, the results are erroneous, significantly modified when material dispersion is considered. To correctly take dispersion into account, one should turn to the $Q$-factor definition, making sure that the stored energy is correctly calculated. In the case of materials with dispersive conductivity such as graphene, an additional energy term must be incorporated [19]; the energy definition of Eq. (8) is then modified to

$$W = \frac{1}{4}\iint_A \varepsilon_0 \left.\frac{\partial\{\omega\varepsilon_r(\omega)\}}{\partial\omega}\right|_{\omega=\omega_0} |\mathbf{E}|^2 \mathrm{d}S + \frac{1}{4}\iint_A \mu_0|\mathbf{H}|^2\mathrm{d}S + \frac{1}{4}\int_{\mathrm{gr}} \left.\frac{\partial \mathrm{Im}\{\sigma_{\mathrm{gr}}(\omega)\}}{\partial\omega}\right|_{\omega=\omega_0} |\mathbf{E}_{\parallel}|^2\mathrm{d}\ell. \tag{12}$$

Using Eq. (12) to calculate the stored energy in the resonator and, consequently, the quality factor, an approximately twofold increase in all $Q_i$, $Q_{\mathrm{res}}$, and $Q_{\mathrm{rad}}$ is observed as Fig. 8 indicates, exclusively due to the dispersive energy introduced by the last term. For highly dispersive resonators though, it must be noted that the independency of the $W/P_{\mathrm{loss}}$ ratio on the size of the integration domain ceases to hold, as evidenced from Eq. (5), which despite being general, actually relates losses to an energy term accounting for stored energy in a medium without dispersion. Since dispersion is not straightforwardly considered in a typical eigenproblem formulation, only the "nondispersive" part of the energy $W$ scales with the integration domain extent (due to the exponential increase in the radiated field as the distance increases), while the dispersive part, mostly originating from the bound field in the vicinity of the dispersive resonator, remains constant. However, in this case of rather weak radiation, $W$ is constant regardless of the integration domain and dispersive $Q_i$, $Q_{\mathrm{res}}$, and $Q_{\mathrm{rad}}$ are correctly calculated.

External and loaded quality factors are calculated following the same approach as in Sec. 3. Coupling loss ($P_e$) is obtained by integrating the outward Poynting vector over the two green dotted lines ($\partial \mathrm{A}_{\mathrm{wg}}$) of Fig. 7(b) and the stored energy $W$ is given by Eq. (12), with Box A′ as the integration domain. The results for the $m = 4$ order mode, depicted for various coupling gaps in Fig. 9(a), agree remarkably with the calculation of the quality factor using the "spectral response" method. Still, if one erroneously neglects graphene dispersion and uses Eq. (8) to calculate the stored energy confirms that the obtained results are in very good agreement with those obtained using the "eigenfrequency" method, yet overall wrong [Fig. 9(b)]. This is further highlighted in Fig. 10, where the spectral response of the $m = 4$ order mode, obtained after driven time-harmonic simulations, is compared with the CMT results of Eq. (10) when using the quality factors calculated from the "eigenmode" method, including the extra energy term (blue thick curve) or the "eigenfrequency" method (red thin curve). It is obvious that the former approach returns the correct $Q$ values, in any of the three cases considered corresponding to overcoupling ($g = 1.54$ $\mu$m), critical coupling ($g = 1.74$ $\mu$m), and undercoupling ($g = 1.94$ $\mu$m) conditions, respectively. We must note here that regardless of the correct incorporation of graphene dispersion, the critical coupling condition is in both cases recovered [cf. Fig. 9], since it actually corresponds to a power loss equality, i.e., $P_e = P_i$. The above remarks are gathered in Table 2, where it is emphatically highlighted that the "eigenfrequency" method cannot be used to calculate any quality factor in highly dispersive systems.

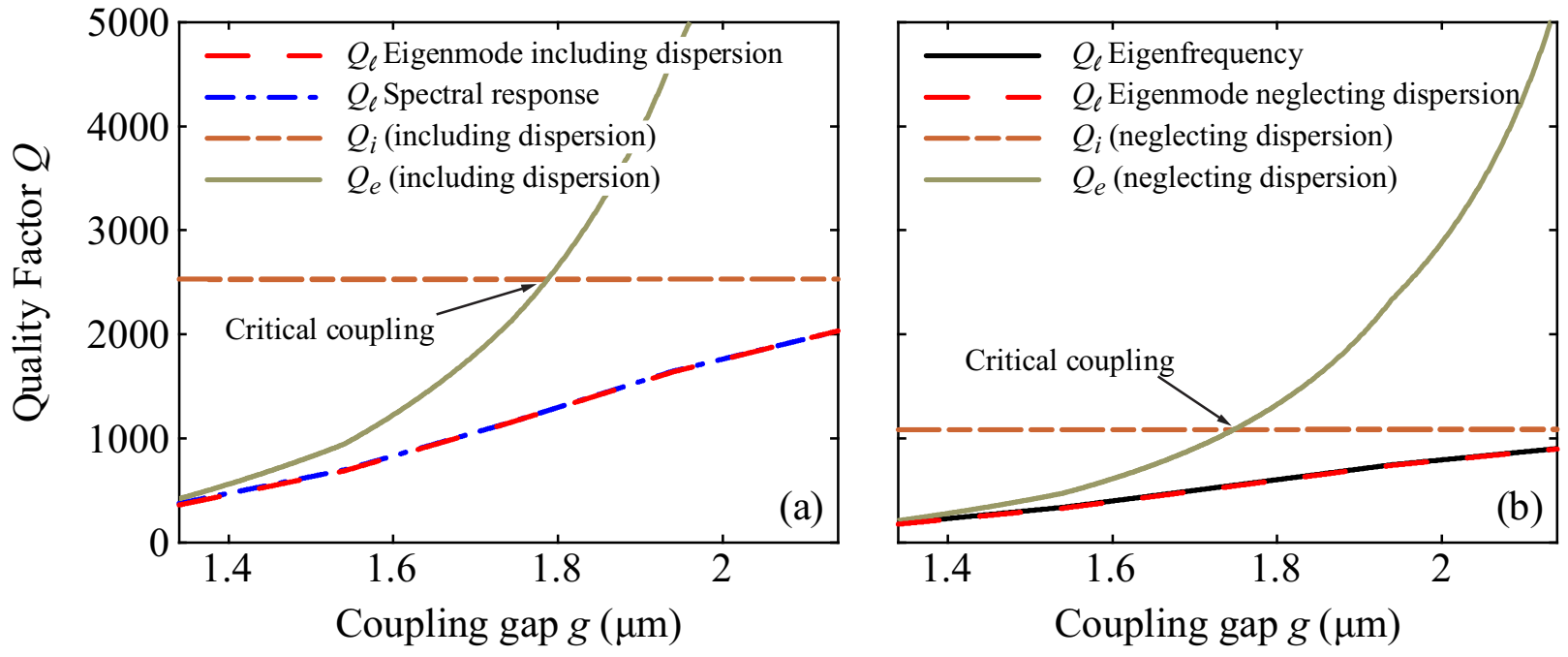

Fig. 9. (a) Loaded $Q$-factor obtained from the eigenmode (red dashed line) and the spectral response (blue dashed-dotted line) when graphene dispersion is included. (b) Loaded $Q$-factor obtained from the eigenfrequency (black solid line) and the eigenmode (red dashed line) when graphene dispersion is erroneously neglected. $Q_i$ and $Q_e$ are also included to identify critical coupling conditions, actually being comparable in both considerations.

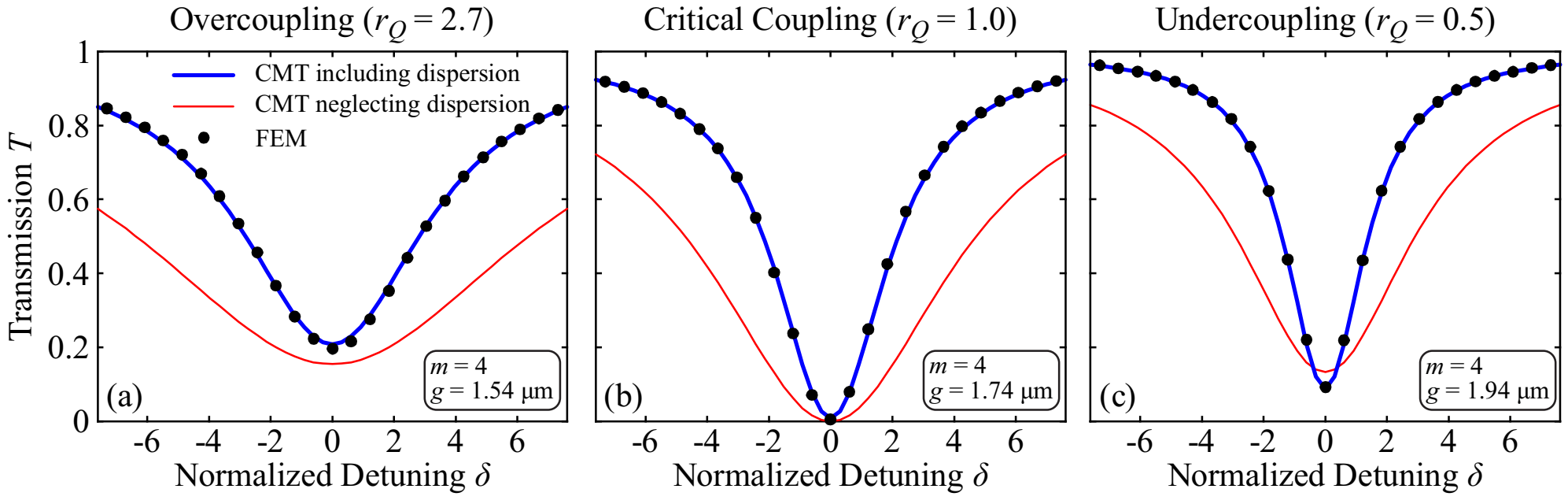

Fig. 10. FEM results (black dots) compared to CMT, with the $Q$-factors calculated including (blue thick curve) or ignoring (red thin curve) graphene dispersion. The correct spectral response is recovered when dispersion is taken into account.

**Table 2. $Q$-factor Calculation Methods in Guided-Wave Systems with Strong Material Dispersion, Significant Resistive, and Negligible Radiation Loss. n/a stands for "not applicable."**

| | Eigenproblem | | Time-harmonic (waveguide fed) | |
|---|---|---|---|---|
| Method | Eigenfrequency | Eigenmode | Field distribution | Spectral response |
| $Q_{\text{res}}$ | ✗ | ✓ | ✓ | n/a |
| $Q_{\text{rad}}$ | ✗ | ✓ | ✓ | n/a |
| $Q_i$ | ✗ | ✓ | ✓ | n/a |
| $Q_e$ | ✗ | ✓ | ✓ using $\mathbf{E}^{\text{sc}}$ | n/a |
| $Q_\ell$ | ✗ | ✓ | ✓ using $\mathbf{E}^{\text{sc}}$ | ✓ |

## 5. Polaritonic rod metasurface at THz frequencies

Thus far, we have investigated the applicability of the most prominent $Q$-factor calculation methods using two indicative guided-wave resonant systems. In what follows, we will study free-space resonant structures. Specifically, we will systematically examine the applicability range of the presented $Q$-factor calculation methods in a dielectric rod meta-atom (low THz regime) and the respective metasurface [Fig. 1(c)], and, subsequently, in a plasmonic core-shell nanoparticle (visible regime) [Fig. 1(d)]. We first examine a dielectric cylindrical resonator [Fig. 11(a)] of radius $R$ in the order of 8 $\mu$m [36]. The dielectric rod is made of lithium tantalate ($LiTaO_3$), which is a high-index, low-loss, and weakly dispersive material in the low THz regime (far below the phonon resonances) [37], where the rod actually resonates (around 2 THz).

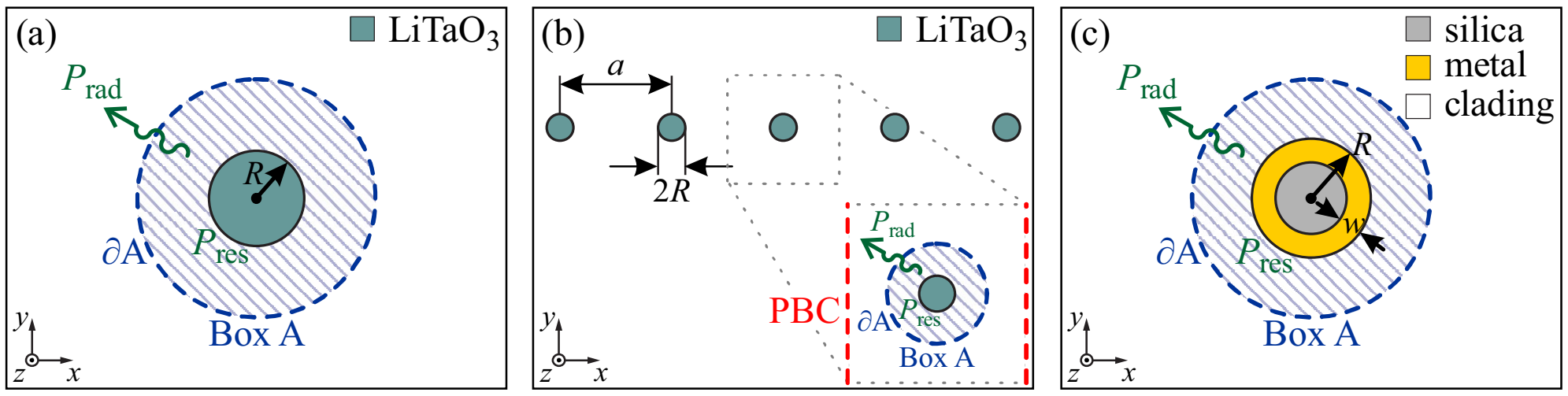

Fig. 11. Schematics of (a) a $LiTaO_3$ microrod of radius $R$, (b) a $LiTaO_3$ metasurface with lattice constant $a$, and (c) a plasmonic core-shell nanoparticle of outer radius $R$ with a silica core and a gold shielding of width $w$. A typical integration domain and the respective power integration limit for applying Eq. (1) are sketched in all structures.

Before assessing the $Q$-factor calculation methods, we note that the concept of external coupling in free-space systems is essentially different from the so far examined guide-wave resonators. Specifically, light couples to the atom from the illumination field (commonly, an appropriately polarized plane wave or Gaussian beam) and thus the distinction between coupled and uncoupled systems is here meaningless. Thus, when free-space resonators are considered, their $Q_{\text{rad}}$ corresponds directly to $Q_e$.

Returning to the $LiTaO_3$ rod [Fig. 11(a)], we examine the first four modes (Mie resonances) of TM polarization ($\mathbf{H} \equiv H_z\hat{\mathbf{z}}$). Keeping the resonance frequency around 2 THz, we find the $TM_{00}$ mode supported for $R = 8$ $\mu$m, $TM_{10}$ for $R = 14$ $\mu$m, $TM_{20}$ for $R = 18$ $\mu$m, and $TM_{01}$ for $R = 20$ $\mu$m. The complex resonance frequency of each mode is determined by solving the respective eigenproblem (iteratively to correctly account for dispersion). The intrinsic $Q$-factors, calculated via the "eigenfrequency" method [black circles in Fig. 12(a)], range from 16 ($TM_{00}$) up to 276 ($TM_{20}$), coinciding with the "eigenmode" method including dispersion (yellow crosses), applied to an arbitrary-radius Box A [Fig. 11(a)], indicating that material dispersion is indeed weak. $Q_{\text{res}}$ lies around 550 for all four modes, implying that radiation loss dominates in the three low-$Q$ modes ($TM_{00/10/01}$). For the low-dispersion system under study, $Q_{\text{rad}}$ is calculated using the "eigenfrequency" method after setting $\varepsilon_r'' = 0$ to momentarily zero out resistive losses. Subsequently, $Q_{\text{res}}$ is found using Eq. (2). One should be particularly carefully when applying the $Q$-factor definition using the eigenmode to calculate $Q_{\text{rad}}$ and $Q_{\text{res}}$ since $W/P_{\text{rad}}$ and $W/P_{\text{res}}$ are not constant with respect to the integration domain dimensions, as $W/P_i$ is [cf. Eq. (5) with $P_e = 0$]. In fact, this method cannot be applied using the eigenmode, especially when $P_{\text{res}}$ and $P_{\text{rad}}$ are of comparable order of magnitude. Nevertheless, one can momentarily neglect losses, calculate $Q_{\text{rad}}$ through Eq. (1) and finally use Eq. (2) to obtain $Q_{\text{res}}$. An alternative approach to calculate $Q_{\text{res}}$ and $Q_{\text{rad}}$ is with the "field distribution" method. The rod can be illuminated with two counter-propagating $x$-polarized plane waves coming from the $y$ and $-y$ directions with appropriate phase difference to correctly excite the respective mode (0 for symmetric and $\pi$ for antisymmetric). The scattered field (i.e., full field minus the two illumination waves) $\mathbf{E}^{\text{sc}}$ should, of course, be used for the calculations. Still, even though $P_{\text{rad}}$ is constant regardless of the integration domain limit $\partial A$, the stored energy $W$ increases with the integration domain extend, encompassing additional energy associated with the radiated field. Thus, the "field distribution" method can and should be only applied for relatively small integration domains to correctly estimate $Q$, taking care to accommodate only the reactive power associated with the near-field of the mode and as little as possible of the propagating radiation.

The "spectral response" method (blue squares) can also be applied by illuminating the rod with an $x$-polarized plane wave propagating along $y$ axis and calculating the absorption cross-section, which exhibits a clear Lorentzian lineshape for each resonance. The absorption cross-section is calculated through

$$C_{\text{abs}} = \frac{\frac{1}{2}\iint_{\text{rod}} \omega\varepsilon_0\varepsilon_r''|\mathbf{E}|^2 \mathrm{d}S}{I_0}, \tag{13}$$

where the integration is exclusively performed inside the lossy $LiTaO_3$; $I_0 = E_0^2/2\eta_0$ is the illumination field intensity and $\eta_0 \approx 120\pi$. The results of the "spectral response" method coincide with the other two, since material dispersion is weak.

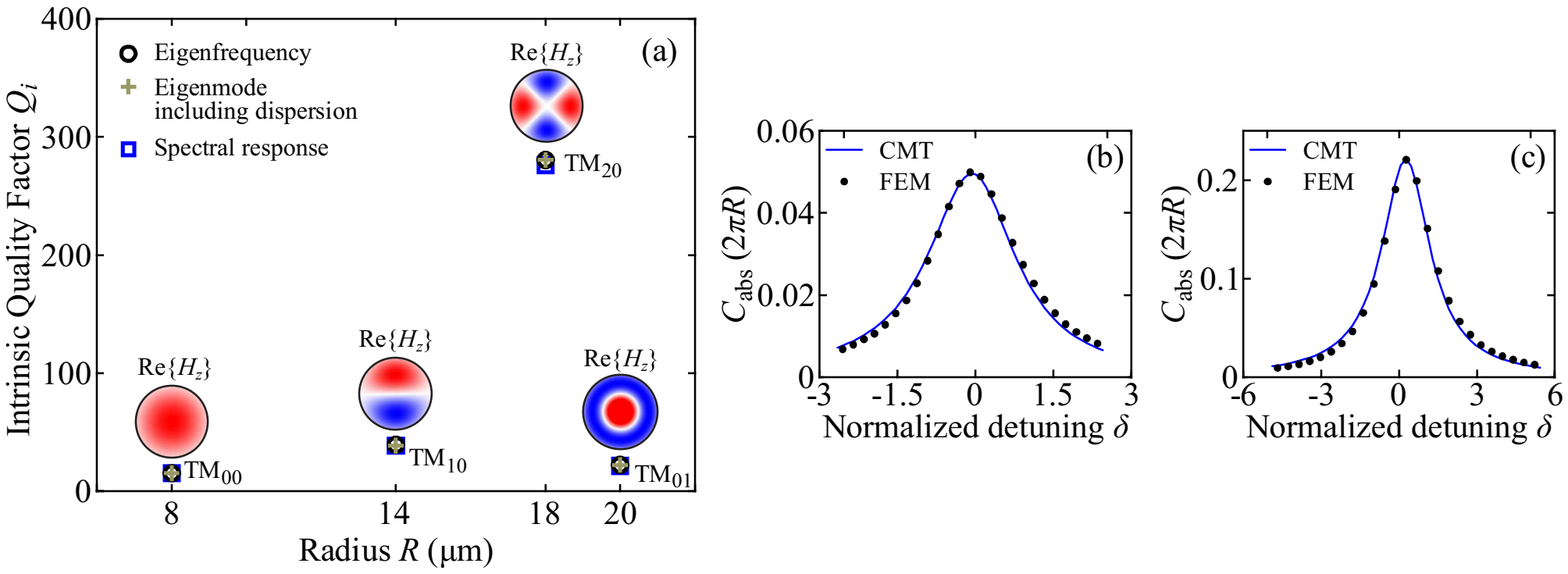

Fig. 12. (a) Intrinsic $Q$-factor for low order Mie resonances (TM polarization, $\mathbf{H} \equiv H_z\hat{\mathbf{z}}$) of the dielectric meta-atom for resonance frequencies around 2 THz. All the $Q$-factor calculation methods coincide. Absorption cross-section calculated using FEM simulations (black dots) and CMT (blue solid curves) for (b) the $TM_{00}$ mode and (c) the $TM_{10}$ mode. FEM-CMT agreement is excellent.

The absorption cross-section can also be recovered using an appropriate CMT framework [38, 39]. In particular, for 2D structures it is given by

$$C_{\text{abs}} = \frac{2(m+1)\lambda}{\pi}\frac{r_Q}{\delta^2 + (1+r_Q)^2}, \tag{14}$$

where $\lambda = \lambda_0/n$, $\delta = 2Q_{\text{rad}}(\omega - \omega_0)/\omega_0$, $r_Q = Q_{\text{rad}}/Q_{\text{res}}$, and $m$ is the azimuthal order of the mode. The results (expressed in units of $2\pi R$) for the first two modes are depicted in Figs. 12(b) and 12(c) (blue solid curves), respectively, and are in complete agreement with the FEM-based calculated absorption cross-section (black dots), Eq. (13).

The dielectric meta-atom examined in this section is commonly used as the fundamental unit cell for realizing all-dielectric metasurfaces, capable of purposefully manipulating light [40]. The correct estimation of the $Q$-factor in this case is equally important. For the simulations, we consider the periodic expansion of our initial meta-atom, computationally modeled by using appropriate periodic boundary conditions (PBCs), with a subwavelength lattice constant $a = 50\ \mu$m to avoid the excitation of higher diffraction orders, Fig. 11(b). The same principles regarding $Q$-factor calculation hold here as well; the "eigenfrequency," the "eigenmode," and the "field distribution" methods give the same $Q$ value. For brevity, in Fig. 13 we only present the metasurface absorption spectrum $A \equiv P_{\text{abs}} = \frac{1}{2}\iint_{\text{rod}} \omega\varepsilon_0\varepsilon_r''|\mathbf{E}|^2\mathrm{d}S$ (black dots), calculated using driven time-harmonic simulations, and the respective CMT calculations $A = 2r_Q/[\delta^2+(1+r_Q)^2]$ (blue solid curve), showing excellent agreement; the respective $Q$-factors are included for reference in Fig. 13 next to the corresponding absorption peak. For the demonstration, we chose absorption over transmission or reflection, since the latter exhibit Fano lineshapes due to the interplay between the resonance and the direct pathway. Although CMT is capable of handling such scenarios given that the individual resonances and the direct pathway are correctly assessed [41], an in-depth examination of the process lies outside the scope of this work. In Table 3 we summarize the applicability range of the $Q$-calculation methods in free-space systems with low dispersion and non-negligible radiation/resistive losses.

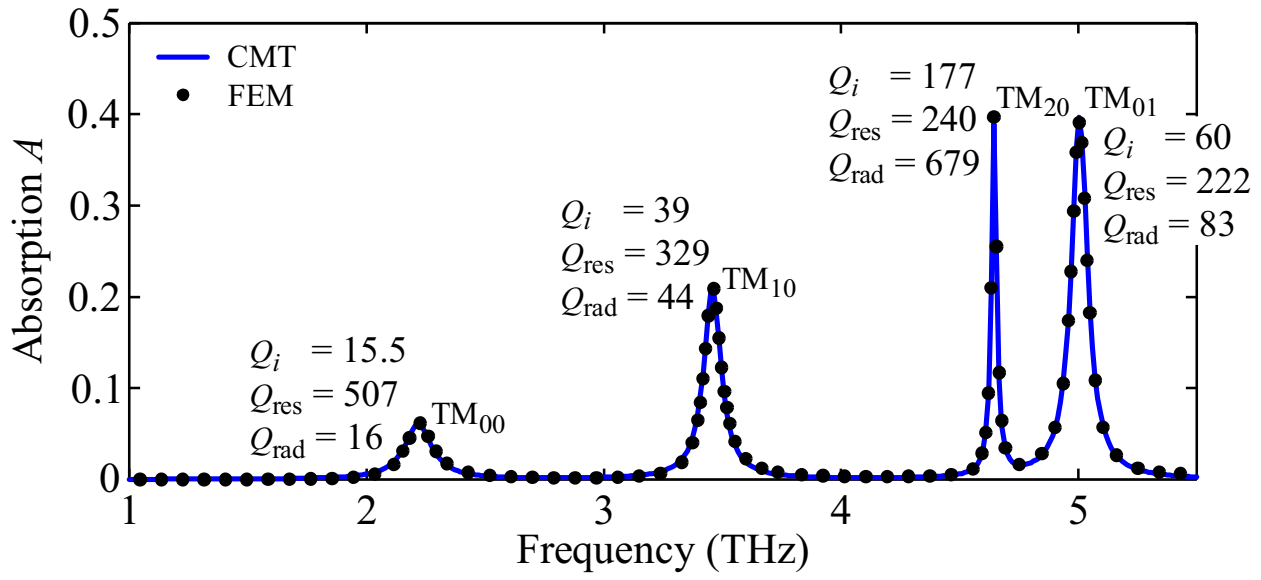

Fig. 13. Dielectric metasurface absorption using FEM simulations (black dots) and CMT (blue solid curve), revealing excellent agreement for all four modes.

**Table 3. $Q$-factor Calculation Methods in Strongly Leaky and Lossy Scatterers with Negligible Dispersion. n/a stands for "not applicable."**

| | Eigenproblem | | Time-harmonic (wave illumination) | |
|---|---|---|---|---|
| Method | Eigenfrequency | Eigenmode | Field distribution | Spectral response |
| $Q_{res}$ | ✓ using Eq. (2) | ✓ using Eq. (2) | ✓* | n/a |
| $Q_{rad}$ | ✓ with $\varepsilon_r'' = 0$ | ✓ with $\varepsilon_r'' = 0$ | ✓* | n/a |
| $Q_i$ | ✓ | ✓ | ✓* | ✓ |

* using $\mathbf{E}^{sc}$ in an appropriately selected integration domain, accommodating only the reactive near fields.

## 6. Plasmonic core-shell nanoparticle at the visible

Finally, we study a plasmonic nanoparticle exhibiting important radiation and ohmic loss as well as high material dispersion. Plasmonic nanoparticles have been used in various applications [42–44] and thus the correct calculation of their resonance frequency and their quality factor is of outmost practical interest. Owing to their plasmonic nature and subwavelength dimensions, the correct calculation of the $Q$-factor is rather challenging and requires considerable attention [22]. To highlight the correct approach to calculate $Q$ using commercially available FEM software, we examine a plasmonic core-shell nanorod, resonating in the visible [45], Fig. 1(d). We choose core-shell structures since they provide an extra degree of freedom in controlling the ratio between ohmic and radiation loss. Specifically, a cylindrical core-shell of outer radius $R$ in the order of 110 nm is examined, consisting of a silica core and a gold shielding of width $w$ in the order of 45 nm, with a resonance wavelength arround 500 nm, Fig. 11(c). The permittivity of gold is given by a standard Drude dispersion model with $\omega_p = 1.26 \times 10^{16}$ rad/s and $\Gamma = 7 \times 10^{13}$ rad/s, chosen over more accurate permittivity models since we are interested in the presence of strong dispersion rather in its specific form. Silica refractive index is 1.4618 [29] in the visible spectrum, and the homogenous cladding has a refractive index of 1.5.

The analysis is focused on four supported modes with relatively high $Q$-factors. Keeping the resonance frequency around 500 nm and the $R/w$ ratio constant at 2.44, we focus on the $TM_{10}$, $TM_{20}$, $TM_{30}$, and $TM_{01}$ modes. $TM_{00}$ is also supported, being though unsuitable for further examination due to its quite low $Q$-factor. The intrinsic $Q$-factors [Fig. 14(a)], obtained using the "eigenfrequency" method (black circles) range from 17 ($\mathrm{TM}_{10}$) up to 440 ($\mathrm{TM}_{01}$), coinciding with the "eigenmode" method (yellow crosses), applied to an arbitrary-radius Box A [Fig. 11(c)] neglecting, at first, material dispersion. Due to their quasi-plasmonic nature, $\mathrm{TM}_{m0}$ modes experience important ohmic loss, in contrast to the quasi-photonic mode $\mathrm{TM}_{01}$ [Fig. 14(b)], while radiation loss is limited for higher radii [Fig. 14(c)]. Regarding the calculation of $Q_{\mathrm{res}}$ and $Q_{\mathrm{rad}}$ using the "field distribution" method (yellow stars), applied after the illumination of the core-shell with two counter-propagating plane waves, one should be cautious, as already highlighted in Sec. 5.

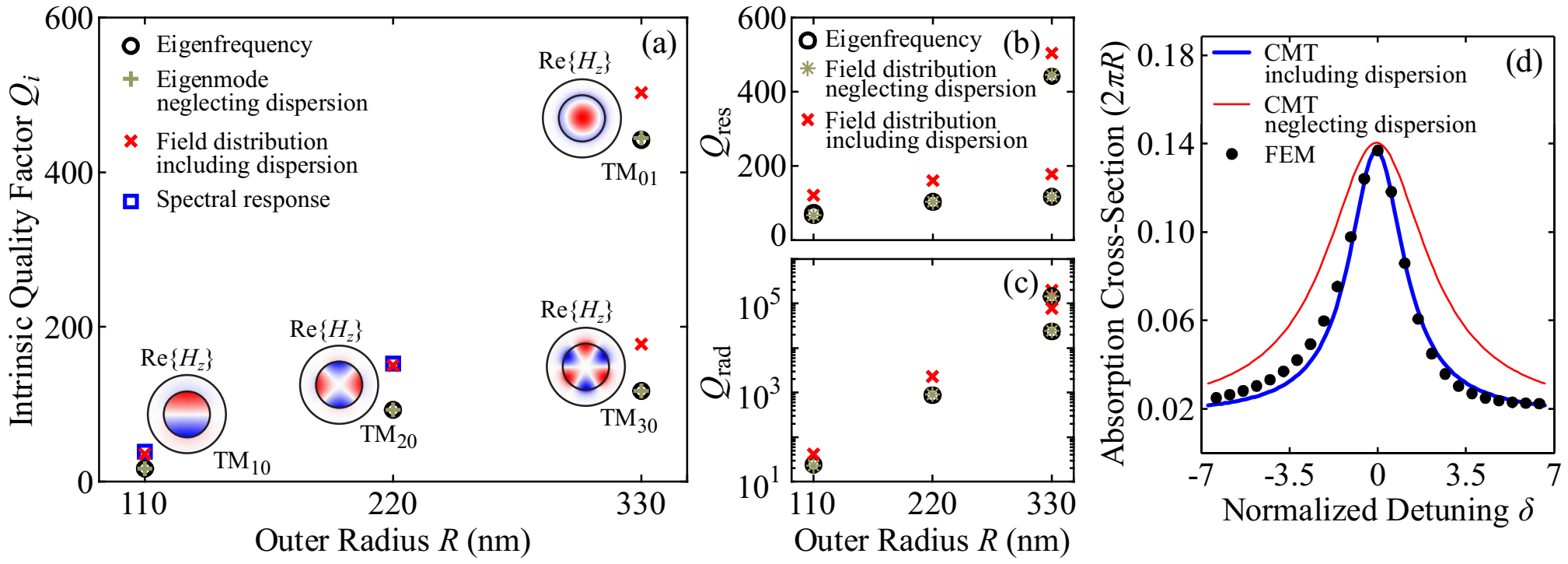

Fig. 14. (a) Intrinsic $Q$-factor for various core-shell modes. "Eigenfrequency" (black circles) and "eigenmode" (yellow crosses) methods coincide but both fail to capture material dispersion. Field distribution (red crosses) and spectral response (blue squares) correctly incorporate dispersion. All modes have resonance wavelengths around 500 nm. $Q_i$ may be carefully decomposed in its (b) $Q_{\mathrm{res}}$ and (c) $Q_{\mathrm{rad}}$ contributions. (d) FEM (black dots) and CMT results using the calculated $Q$-factors including (blue thick curve) or neglecting (red thin curve) material dispersion. The correct absorption cross-section spectral response is recovered when dispersion is correctly taken into account.

Taking dispersion into account (both in gold and silica) we find that it induces significant variations in the obtained $Q$-factors, mainly for the quasi-plasmonic $\mathrm{TM}_{m0}$ modes (approximately 60% increase) due to their strong interaction with gold, in contrast to the quasi-photonic mode (exhibiting a slight 10% increase). To determine $Q_i$ (and subsequently $Q_{\mathrm{res}}$ and $Q_{\mathrm{rad}}$) only the "field distribution" method can be applied since, as already mentioned in Sec. 4, the ratio $W/P_i$ depends on the size of the integration domain when the dispersive energy expression [Eq. (8)] is employed. The results, depicted with red crosses in Fig. 14(a) are verified by comparison with the "spectral response" method (blue squares), after calculating the FWHM of the absorption cross-section spectrum [Eq. (13)]. Unfortunately, for $R = 330$ nm, several low-$Q$ leaky plasmonic modes are supported at the gold-cladding interface, and thus a clear Lorentzian peak cannot be recovered. Additionally, $Q_{\mathrm{res}}$ and $Q_{\mathrm{rad}}$ cannot be directly validated using the "spectral response" method, unless we resort in ancillary spectral sweeps. Thus, we rely on the correct calculation of $Q_i$ at an appropriate integration domain for subsequently calculating $Q_{\mathrm{res}}$ and $Q_{\mathrm{rad}}$.

To test this hypothesis, we substitute the obtained $Q$-factors in the respective CMT framework [Eq. (14)]. The results for the $\mathrm{TM}_{10}$ mode (expressed in units of $2\pi R$) using the $Q$-factor calculations with (blue thick curve) or without (red thin curve) considering dispersion are depicted in Fig. 14(d) and compared with the FEM-based absorption cross-section [Eq. (13)] where dispersion is explicitly taken into account (black dots). Although FEM and CMT do not exhibit the excellent concurrence of the previous examples (attributed to the complex absorption cross-section spectrum due to neighbouring resonances with extremely low $Q$), it is clear that the correct incorporation of dispersion when calculating the $Q$-factor is crucial (let us note here that both CMT curves have been shifted upwards to account for the absorption background of the neighbouring, low-$Q$ mode contributions). These discussions are summarized in Table 4, where it is emphatically highlighted that neither the "eigenfrequency" nor the "eigenmode" methods can be used to calculate any quality factor in leaky and lossy scatterers with high dispersion. To overcome this barrier, other alternatives should be adopted such as different formulations of the eigenvalue problem to correctly incorporate dispersion [35], or possibly follow a route by regularizing the QNM to eliminate the exponential divergence [26].

**Table 4. $Q$-factor Calculation Methods in Leaky and Lossy Scatterers with High Dispersion. n/a stands for "not applicable."**

| | Eigenproblem | | Time-harmonic (wave illumination) | |
|---|---|---|---|---|
| Method | Eigenfrequency | Eigenmode | Field distribution | Spectral response |
| $Q_{\mathrm{res}}$ | ✗ | ✗ | ✓* | n/a |
| $Q_{\mathrm{rad}}$ | ✗ | ✗ | ✓* | n/a |
| $Q_i$ | ✗ | ✗ | ✓* | ✓ |

* using $\mathbf{E}^{\mathrm{sc}}$ in an appropriately selected integration domain, accommodating only the reactive near fields.

## 7. Conclusion

To recapitulate, we have identified and organized the commonly used methods for calculating the quality factor in guided-wave and free-space resonant systems based on spectral simulations with commercially available, general purpose software. We have thoroughly assessed the applicability range of these methods through judiciously selected examples covering different material systems and frequency regimes from the far-infrared to the visible, and have highlighted the underlying physical and numerical reasons limiting the applicability of each one.

We have shown that all methods produce correct results in conventional photonic systems exhibiting moderate to low material dispersion, tolerable ohmic loss, and potentially significant radiation leakage. However, when it comes to state-of-the-art systems (plasmonics, 2D materials, etc.) one should be extra careful, mainly due to the physical complication of strong material dispersion. By studying structures exhibiting different levels of dispersion, ohmic loss, and radiation leakage, we have highlighted the pitfalls that arise and identified the correct approaches for obtaining sound results, despite the fact that most of the commonly considered calculations fail. More specifically: (a) The "eigenfrequency" method fails to account for material dispersion, leading to important deviations in the calculated $Q$-factors when dispersion is strong. The error is proportional to the impact of dispersion on the stored energy. (b) In the "eigenmode" method we can take care in correctly calculating the energy in the general dispersive case. As a result, the method can be applied to study systems with strong dispersion but only when radiation leakage is low and the corresponding exponential divergence of the eigenmode is not pronounced. (c) The "field distribution" method, on the other hand, does not suffer from such limitations; however, when applied in strongly leaky systems, caution is required to avoid accounting for energy associated with the propagating radiated wave. (d) Finally, the "spectral response" method, always returns correct results provided that the resonance is well separated in frequency from neighboring resonant processes. For easy reference, the above-discussed remarks are illustrated in Fig. 15, visually summarizing the key findings of our work.

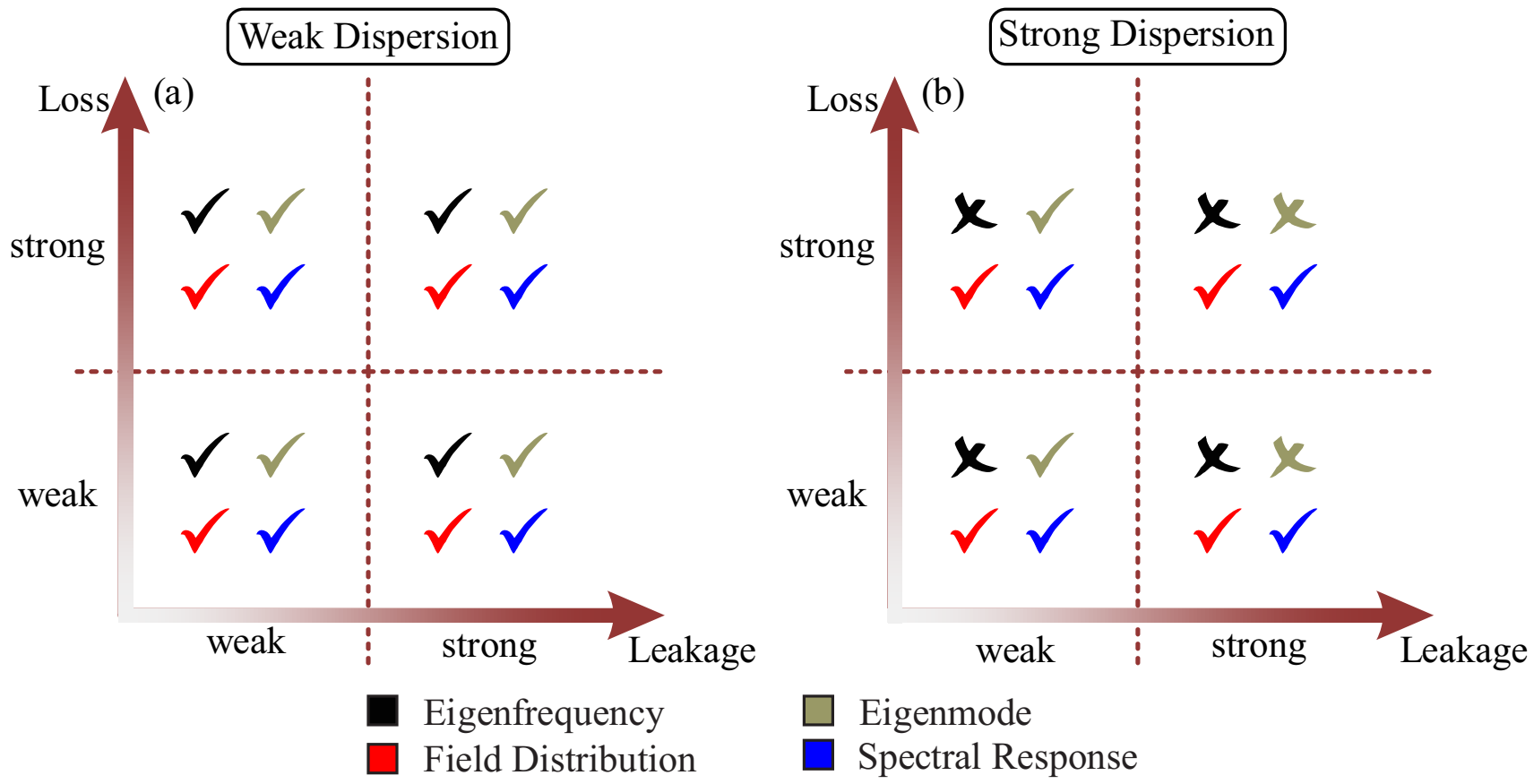

Fig. 15. Applicability of $Q$-factor calculation methods, schematically imprinted in the Leakage-Loss "plane", when dispersion is (a) weak and (b) strong.

## Acknowledgment

The authors acknowledge the tangible contribution of Mr. Vasileios Ataloglou on the preliminary simulations of this work. This research was co-financed by Greece and the European Union (European Social Fund-ESF) through the Operational Program "Human Resources Development, Education and Lifelong Learning 2014-2020" in the context of the project "Nonlinear phenomena in graphene-comprising resonators" (MIS 5004717).